\documentclass{pasa}%

\title[A globular cluster in And I]{The faint globular cluster in the dwarf galaxy Andromeda I}
\author[Caldwell et al.]{Nelson Caldwell$^1$, Jay Strader$^2$, David J.~Sand$^{3,4}$, Beth Willman$^5$, Anil C. Seth$^6$\\
\affil{$^1$Harvard-Smithsonian Center for Astrophysics, 60 Garden Street, Cambridge, MA 02138, USA; caldwell@cfa.harvard.edu}
\affil{$^2$Center for Data Intensive and Time Domain Astronomy, Department of Physics and Astronomy, Michigan State University, East Lansing, MI 48824, USA; strader@pa.msu.edu}
\affil{$^3$Department of Physics and Astronomy, Texas Tech University, Box 41051, Lubbock, TX 79409, USA}
\affil{$^4$Steward Observatory, University of Arizona, 933 North Cherry Avenue, Tucson, AZ, 85721, USA}
\affil{$^5$LSST and Steward Observatory, 933 North Cherry Avenue, Tucson, AZ 85721, USA}
\affil{$^6$Department of Physics and Astronomy, University of Utah, 115 South 1400 East, Salt Lake City, Utah 84112, USA}}

\jid{PASA}
\doi{10.1017/pas.\the\year.xxx}
\jyear{\the\year}

\usepackage{aas_macros}
\usepackage{hyperref} 
\hypersetup{colorlinks,citecolor=blue,linkcolor=blue,urlcolor=blue}

\hypersetup{draft}
\begin{document}

%\begin{frontmatter}

\begin{abstract}
  Observations of globular clusters in dwarf galaxies can be used to study a variety of topics, including the structure of dark matter halos and the history of vigorous star formation
  in low-mass galaxies. We report on the properties of the faint globular cluster ($M_V \sim -3.4$) in the M31 dwarf galaxy Andromeda I. This object adds to the growing population of low-luminosity Local Group galaxies that host single globular clusters.
 \end{abstract}

%\end{frontmatter}

%-------------------------------------------------------------------
\maketitle
\section{Introduction}

Globular clusters (GCs) have long been used as luminous tracers of the star formation histories, chemical enrichment, and gravitational potential of their host galaxies, especially 
in the stellar halos where the density of stars is low (Brodie \& Strader 2006). GCs in dwarf galaxies are of particular interest; for example, in recent years they have been used to provide evidence that some dark matter halos have central cores rather than cusps (e.g., Goerdt et al.~2006) and to constrain the extreme mass loss from GCs predicted in some models
of the formation of multiple stellar populations (Larsen et al.~2012). The presence of GCs also hints at active early star formation in dwarfs, as measurements in the local universe suggest that the fraction of stars formed in bound clusters and the maximum mass of star clusters correlates strongly with the star formation rate density (e.g., Goddard et al.~2010; Cook et al.~2012; Kruijssen 2012; Johnson et al.~2017).

Most dwarfs in the Local Group do not host GCs, especially among the less massive galaxies. The least-luminous Local Group galaxy known to have a GC is the recently discovered dwarf
Eridanus~II. This galaxy has $M_V = -7.1$ and hosts a GC with $M_V = -3.5$, which by itself makes up 4\% of the luminosity of its host (Koposov et al.~2015; Crnojevi\'{c} et al.~2016a). 
Contenta et al.~(2017) and Amorisco (2017) argue that the survival of this GC against dynamical friction, despite its projected location close to the galaxy center (offset by only $\sim 45$ pc), is evidence for a cored dark matter halo in Eridanus~II: if a standard cuspy halo were present the GC would have been very unlikely to have survived to the present day. Amorisco (2017) makes a similar argument for the M31 satellite Andromeda~XXV, which hosts a faint extended star cluster near its center, though the precise offset is not well-determined as the galaxy center itself is poorly constrained (Cusano et al.~2016). These conclusions are important as it is unclear whether galaxies with such low luminosities can effectively transform cuspy dark matter profiles to cored ones through central bursts of star formation (e.g., El-Badry et al.~2016; Read et al.~2016), and the presence of cores could provide evidence for self-interacting dark matter rather than standard cold dark matter (e.g., Robles et al.~2017).

Here we present a discussion of another faint Local Group galaxy with a GC: the M31 satellite Andromeda~I (And~I). The existence of this GC, which we term And~I--GC1, was briefly noted in a conference proceeding many years ago (Grebel et al.~2000) but the basic data of the cluster have not been discussed in the subsequent literature, and it has not been included in lists of Local Group dwarfs that host GCs (e.g., Mackey 2015).

\section{Background, Data Reduction, and Analysis}

Grebel et al.~(2000) noted the presence of a faint GC in the archival \emph{Hubble Space Telescope} ($HST$)/WFPC2 imaging of And~I published in da Costa et al.~(1996). This GC is located at a J2000 degree position (R.A., Dec.) = (11.42881, +38.03162), calculated from our own images discussed below. This position is located $\sim 207$ pc from the center of And~I (we use a distance of 745 kpc; McConnachie~2012). As the half-light radius of And~I is $672\pm69$ pc (McConnachie 2012), the GC sits well within the central regions of the galaxy.
New, deep $HST$/ACS observations of And~I have been obtained as part of program GO-13739 (Skillman et al.~2017), which fortunately include the GC. These data comprise a total exposure time of approximately 29 ksec in $F475W$ and 23 ksec in $F814W$. We retrieved the individual CTE-corrected exposures from the $HST$ archive then processed the data using {\tt DrizzlePac}. The relative exposure offsets were determined using {\tt TweakReg} and then drizzled together with {\tt AstroDrizzle} to a pixel scale of 0.03''.

Using an isophote fitting program, we determined integrated magnitudes of the GC out to a radius of 3'' (100 pixels or 11 pc), the radius at which light from the GC could not be distinguished from the background. As the cluster is faint
and there are a few likely contaminating stars in the field, we masked pixels $4\sigma$ above the local isophote. The integrated Vega magnitudes of the cluster are $F475W=21.7\pm0.1$ and $F814W=20.3\pm0.1$. These photometric uncertainties represent only the shot noise in the masked profile. If the bright stars are not masked, the cluster is brighter by about 0.5 mag
in $F814W$ and 0.3 mag in $F475W$, which should be taken as a qualitative estimate of the systematic uncertainties in the integrated photometry. We note that even in the case of perfect
measurements, there are substantial stochastic fluctuations in the absolute magnitudes and colors of low-mass clusters due to sparse sampling of the stellar mass function and the short lifetimes of bright post-main sequence stars (e.g., Fouesneau et al.~2014).

To reference these measurements to the commonly-used $V$ band, we used Padova single stellar population models (Bressan et al.~2012) for a 13 Gyr population to convert $F475W$ to $V$, finding $V\sim 21.1$, equivalent to $M_V \sim -3.4$ (--3.7 if the bright stars are not masked). The luminosity of And~I GC would rank among the lowest of confirmed M31 GCs, though most GC searches have not been complete to this depth (e.g., Huxor et al.~2014).

To empirically derive a metallicity for the cluster, we corrected the magnitudes for foreground reddening using the maps of Schlafly \& Finkbeiner (2011) and then converted them to AB $g$ and $i$, again using the Padova models. These values are $i_0 = 20.7\pm0.1$ and $(g-i)_0 = 0.70\pm0.11$. As a check on this value, we also calculated integrated colors within the half-light radius (see below), for which we might expect the relative effect of contaminating stars to be lower. This value is $(g-i)_0 = 0.73$, well within the uncertainty on the total color.
Peacock et al.~(2011) publish observed $(g-r)$ and $(r-i)$ color vs.~[Fe/H] plots for M31 GCs; we combine these data to estimate a rough conversion between  $(g-i)$ color and metallicity.
Using the value $(g-i)_0 = 0.70$, we find an estimated photometric metallicity of [Fe/H] = --$1.5\pm0.4$, suggesting the GC is metal-poor, but not remarkably so: its metallicity is typical of the typical of the metal-poor GC populations in the Milky Way and M31. This metallicity is also consistent with the overall low metallicity of And~I itself ([Fe/H] $\sim -1.5$; Kalirai et al.~2010).

We also estimated a rough half-light radius via integrated light, finding a value of $4.2\pm0.4$ pc (the individual $F475W$ and $F814W$ measurements were 4.3 and 4.1 pc, respectively).
This implies a half-mass relaxation time of $\sim 360$ Myr assuming a mean stellar mass of $0.6 M_{\odot}$ (Koposov et al.~2007). The evaporation time will be approximately 10 times longer than the relaxation time, though the actual mass loss rate depends on the current stellar mass function (including remnants), the true galactocentric distance, and the mass of the dark halo (Gieles et al.~2011). In any case, it is clear that the least massive GCs in dwarfs are susceptible to evaporation on timescales of Gyrs, suggesting that clusters like the one in And I are likely to be dissolving.

It is worth considering whether this GC could be an M31 object projected onto And~I, as the dwarf sits at a projected radius of 45 kpc from M31. The surface density profile of M31 GCs from Huxor et al.~(2011) predicts 0.004 GCs per kpc$^{2}$ at the projected distance of And~I. Since the area within the half-light radius itself is $\sim 1.4$ kpc$^{2}$, the expected number of contaminant GCs from M31 is about 0.006. Even if we conservatively increase this estimate by about 50\% to account for incompleteness of the GC search for objects with $M_V > -6$, the predicted number of M31 GCs within the half-light radius of And~I is still $< 0.01$. Given that the GC is much closer to the galaxy center than the half-light radius, this calculation shows the GC is very unlikely to be an M31 interloper, though of course it would be desirable to obtain a radial velocity.

\section{Discussion}

As noted in the Introduction, GCs are rare among Local Group galaxies with low stellar masses, and And~I (with $M_V =$ --11.7; McConnachie 2012) is among the most extreme examples---only Eridanus~II ($M_V = -7.1$) and Andromeda~XXV ($M_V = -9.7$) have lower luminosities and still host a GC. Here we place this galaxy and star cluster in the larger context  of low-mass galaxies with GCs.

The best-studied GC system of a low-mass galaxy in the Local Group is that of the Fornax dwarf spheroidal ($M_V =$ --13.4; McConnachie 2012), which has five clusters with absolute magnitudes ranging from $M_V$ = --5.1 to --8.1 (Strader et al.~2003). One of these GCs, with $M_V =$ --7.3, is more metal-rich than the others and is located near the center of the galaxy, and hence is sometimes cited as a candidate nuclear star cluster, though its radial velocity appears offset from the field stars in the galaxy center (Hendricks et al.~2016). The dynamical friction timescale for the Fornax GCs is less than a Hubble time, so the survival of its GC system is a puzzle; one proposed solution is that the dark matter distribution in the Fornax dSph is cored rather than cusped, leading to a longer inspiral time (Goerdt et al.~2006).

Considering other M31 satellite galaxies: besides the faint And~I and Andromeda~XXV GCs, more luminous clusters were proposed to be affiliated with And XVII ($M_V = -8.7$) by Irwin et al.~(2008), but kinematic studies have shown these GCs are unlikely to be bound to the dwarf (Veljanoski et al.~2014).

The Local Group dwarf irregular WLM hosts one massive GC (Stephens et al.~2006); while WLM is more luminous than the Fornax dwarf spheroidal, its stellar mass is lower, such that this GC makes up at least 5\% of the total mass of the galaxy, and an astonishing $\sim 25$\% of the stars with [Fe/H] $< -2$ (Larsen et al.~2014).

Beyond the Local Group, a number of galaxies in the luminosity range $-10 \lesssim M_V \lesssim -11.5$ likely host at least one GC (e.g., Georgiev et al.~2010; Da Costa et al.~2009). While a firm association between candidate GCs and host galaxies has not been confirmed via spectroscopy for most of these objects, in many cases the GCs are definitively identified via  $HST$ imaging, and it is reasonable to assume that most of these proposed associations are correct. Noting a few of the extremes among this group: the galaxy KK27 ($M_V = -10.1$), with two candidate GCs, is currently the lowest-luminosity galaxy known outside the Local Group with GCs (Georgiev et al.~2010). The IKN galaxy, in the M81 group, is slightly more luminous ($M_V = -11.5$) but has five GCs (the most luminous with $M_V = -8.5$), which together make up at least 10\% of the total stellar mass of this galaxy (Tudorica et al.~2015; Larsen et al.~2014).

Overall, we conclude that, while uncommon, GC systems in dwarf galaxies fainter than $M_V \sim -12$ are not exceptional. Deeper observations of Local Group dwarfs, and the ongoing discovery of low-mass galaxies beyond the Local Group (e.g., Crnojevi\'{c} et al.~2016b), are likely to reveal less massive galaxies with GCs, and the properties of these GCs may offer compelling constraints on the structure of the dark matter halos in these galaxies (Contenta et al.~2017; Amarisco 2017).

   \begin{figure}
   \centering
\includegraphics[scale=0.16]{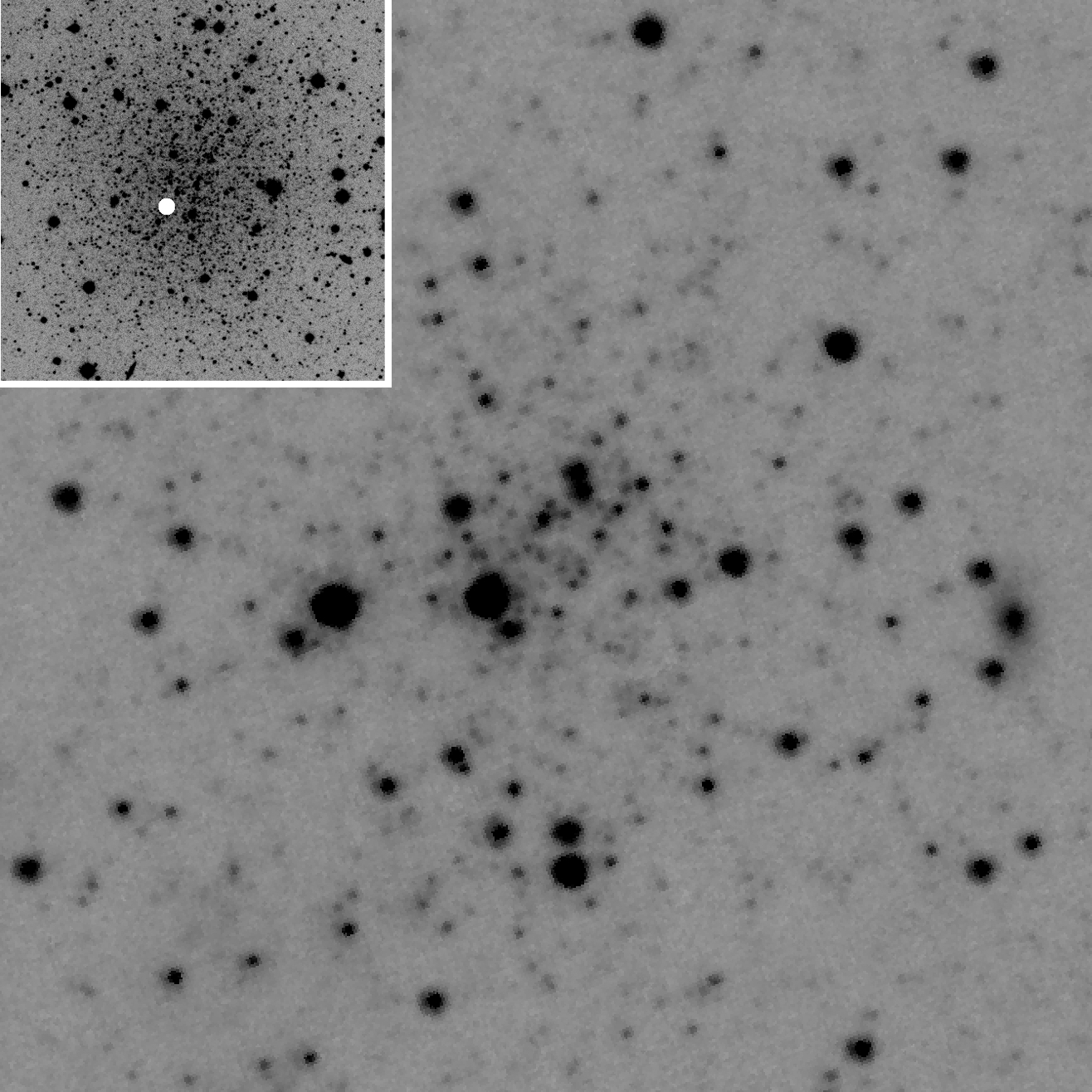}
   \caption{A log-scaled $HST$/ACS $F814W$ image of And~I--GC1, 10'' (38 pc) on a side.  An insert shows the entire And I galaxy, from a ground-based image. This has a width of 7.2' (1.6 kpc). The location of the GC within the dSph is indicated by the white spot.}
              \label{FigGam}%
    \end{figure}

\begin{acknowledgements}
JS acknowledges support from NSF grant AST-1514763.  DJS acknowledges support from NSF grant AST-1412504. This work was partially performed at the Aspen Center for Physics, which is supported by National Science Foundation grant PHY-1066293. 
 \end{acknowledgements}

\bibliographystyle{pasa-mnras}

\begin{thebibliography}{}

\bibitem[Bressan et al.(2012)]{2012MNRAS.427..127B} Bressan, A., Marigo, P., Girardi, L., Salasnich, B., Dal Cero, C., Rubele, S., \& Nanni, A.\ 2012, \mnras, 427, 127 
\bibitem[Brodie \& Strader(2006)]{2006ARA&A..44..193B} Brodie, J.~P., \& Strader, J.\ 2006, \araa, 44, 193 
\bibitem[Conn et al.(2012)]{2012ApJ...758...11C} Conn, A.~R., et al.\ 2012, \apj, 758, 11 
\bibitem[Contenta et al.(2017)]{2017arXiv170501820C} Contenta, F., Balbinot, E., Petts, J.~A., et al.\ 2017, MNRAS, submitted (arXiv:1705.01820)
\bibitem[Cook et al.(2012)]{2012ApJ...751..100C} Cook, D.~O., et al.\ 2012, \apj, 751, 100 
\bibitem[Crnojevi{\'c} et al.(2016a)]{2016ApJ...824L..14C} Crnojevi{\'c}, D., Sand, D.~J., Zaritsky, D., et al.\ 2016, \apjl, 824, L14 
\bibitem[Crnojevi{\'c} et al.(2016b)]{2016ApJ...823...19C} Crnojevi{\'c}, D., Sand, D.~J., Spekkens, K., et al.\ 2016, \apj, 823, 19 
\bibitem[Cusano et al.(2016)]{2016ApJ...829...26C} Cusano, F., Garofalo, A., Clementini, G., et al.\ 2016, \apj, 829, 26 
\bibitem[Da Costa et al.(2009)]{2009AJ....137.4361D} Da Costa, G.~S., Grebel, E.~K., Jerjen, H., Rejkuba, M., \& Sharina, M.~E.\ 2009, \aj, 137, 4361 
\bibitem[Da Costa et al.(1996)]{1996AJ....112.2576D} Da Costa, G.~S., Armandroff, T.~E., Caldwell, N., \& Seitzer, P.\ 1996, \aj, 112, 2576 
\bibitem[El-Badry et al.(2016)]{2016ApJ...820..131E} El-Badry, K., Wetzel, A., Geha, M., et al.\ 2016, \apj, 820, 131 
\bibitem[Fouesneau et al.(2014)]{2014ApJ...786..117F} Fouesneau, M., et al.\ 2014, \apj, 786, 117 
\bibitem[Georgiev et al.(2010)]{2010MNRAS.406.1967G} Georgiev, I.~Y., Puzia, T.~H., Goudfrooij, P., \& Hilker, M.\ 2010, \mnras, 406, 1967
\bibitem[Gieles et al.(2011)]{2011MNRAS.413.2509G} Gieles, M., Heggie, D.~C., \& Zhao, H.\ 2011, \mnras, 413, 2509 
\bibitem[Goddard et al.(2010)]{2010MNRAS.405..857G} Goddard, Q.~E., Bastian, N., \& Kennicutt, R.~C.\ 2010, \mnras, 405, 857 
\bibitem[Goerdt et al.(2006)]{2006MNRAS.368.1073G} Goerdt, T., Moore, B., Read, J.~I., Stadel, J., \& Zemp, M.\ 2006, \mnras, 368, 1073 
\bibitem[Grebel et al.(2000)]{2000AGM....17..P61G} Grebel, E.~K., Dolphin, A.~E., \& Guhathakurta, P.\ 2000, Astronomische Gesellschaft Meeting Abstracts, 17	
\bibitem[Hendricks et al.(2016)]{2016A&A...585A..86H} Hendricks, B., Boeche, C., Johnson, C.~I., Frank, M.~J., Koch, A., Mateo, M., \& Bailey, J.~I.\ 2016, \aap, 585, A86  
\bibitem[Huxor et al.(2011)]{2011MNRAS.414..770H} Huxor, A.~P., et al.\ 2011, \mnras, 414, 770 
\bibitem[Huxor et al.(2014)]{2014MNRAS.442.2165H} Huxor, A.~P., Mackey, A.~D., Ferguson, A.~M.~N., et al.\ 2014, \mnras, 442, 2165 
\bibitem[Irwin et al.(2008)]{2008ApJ...676L..17I} Irwin, M.~J., Ferguson, A.~M.~N., Huxor, A.~P., Tanvir, N.~R., Ibata, R.~A., \& Lewis, G.~F.\ 2008, \apjl, 676, L17 
\bibitem[Johnson et al.(2017)]{2017ApJ...839...78J} Johnson, L.~C., Seth, A.~C., Dalcanton, J.~J., et al.\ 2017, \apj, 839, 78 
\bibitem[Kalirai et al.(2010)]{2010ApJ...711..671K} Kalirai, J.~S., et al.\ 2010, \apj, 711, 671 
\bibitem[Koposov et al.(2007)]{2007ApJ...669..337K} Koposov, S., et al.\ 2007, \apj, 669, 337 
\bibitem[Kruijssen(2012)]{2012MNRAS.426.3008K} Kruijssen, J.~M.~D.\ 2012, \mnras, 426, 3008 
\bibitem[Larsen et al.(2014)]{2014A&A...565A..98L} Larsen, S.~S., Brodie, J.~P., Forbes, D.~A., \& Strader, J.\ 2014, \aap, 565, A98 
\bibitem[Larsen et al.(2012)]{2012A&A...544L..14L} Larsen, S.~S., Strader, J., \& Brodie, J.~P.\ 2012, \aap, 544, L14 
\bibitem[Mackey(2015)]{2015book} Mackey, A.~D.\ 2015, In Lessons from the Local Group, ed. K.~Freeman et al. (Cham, Switzerland: Springer), 215
\bibitem[McConnachie(2012)]{2012AJ....144....4M} McConnachie, A.~W.\ 2012, \aj, 144, 4 
\bibitem[McConnachie \& Irwin(2006)]{2006MNRAS.365.1263M} McConnachie, A.~W., \& Irwin, M.~J.\ 2006, \mnras, 365, 1263 
\bibitem[Peacock et al.(2011)]{2011ApJ...737....5P} Peacock, M.~B., Zepf, S.~E., Maccarone, T.~J., \& Kundu, A.\ 2011, \apj, 737, 5 
\bibitem[Read et al.(2016)]{2016MNRAS.459.2573R} Read, J.~I., Agertz, O., \& Collins, M.~L.~M.\ 2016, \mnras, 459, 2573 
\bibitem[Robles et al.(2017)]{2017arXiv170607514R} Robles, V.~H., Bullock, J.~S., Elbert, O.~D., et al.\ 2017, MNRAS, submitted (arXiv:1706.07514)
\bibitem[Schlafly \& Finkbeiner(2011)]{2011ApJ...737..103S} Schlafly, E.~F., \& Finkbeiner, D.~P.\ 2011, \apj, 737, 103 
\bibitem[Skillman et al.(2017)]{2017ApJ...837..102S} Skillman, E.~D., Monelli, M., Weisz, D.~R., et al.\ 2017, \apj, 837, 102 
\bibitem[Stephens et al.(2006)]{2006AJ....131.1426S} Stephens, A.~W., Catelan, M., \& Contreras, R.~P.\ 2006, \aj, 131, 1426 
\bibitem[Strader et al.(2003)]{2003AJ....125.1291S} Strader, J., Brodie, J.~P., Forbes, D.~A., Beasley, M.~A., \& Huchra, J.~P.\ 2003, \aj, 125, 1291 	 
\bibitem[Tudorica et al.(2015)]{2015A&A...581A..84T} Tudorica, A., Georgiev, I.~Y., \& Chies-Santos, A.~L.\ 2015, \aap, 581, A84 
\bibitem[Veljanoski et al.(2014)]{2014MNRAS.442.2929V} Veljanoski, J., et al.\ 2014, \mnras, 442, 2929 

\end{thebibliography}

\end{document}